\documentclass[preprint2]{aastex}
\usepackage {amsmath}
\usepackage {amssymb}
\usepackage {amsfonts}
\usepackage {amsthm}
\usepackage {mathrsfs}
\usepackage {natbib}
\usepackage {latexsym}
\usepackage {graphicx}
\usepackage {dsfont}
\usepackage {times}
\usepackage {txfonts}
\usepackage {rotating}
\usepackage {wasysym}
\usepackage {multirow}
\usepackage {hhline}
\usepackage {hyperref}
\usepackage {color}
\usepackage {bm}
\usepackage{appendix}
\usepackage{acronym}
\usepackage {url}
\usepackage{subfigure}
\usepackage[normalem]{ulem}   

\hypersetup{
  colorlinks=true,        
  linkcolor=blue,         
  citecolor=cyan,         
}

\slugcomment{Target Astrophysical Journal}

\shorttitle{Multimessenger astronomy}
\shortauthors{Fan et al.}

\begin{document}


\title{Multimessenger astronomy}


\author{XiLong~Fan$^{1,}$\altaffilmark{3}, Martin Hendry$^{2}$}
\affil{1. School of Physics and Electronics Information, Hubei University of Education, 430205 Wuhan, China,\\
2. SUPA, School of Physics and Astronomy, University of Glasgow, Glasgow, G12 8QQ, United Kingdom}


\altaffiltext{3}{Xilong.Fan@glasgow.ac.uk}



\begin{abstract}
In this paper we provide a short overview of the scope and strong future potential of a multi-messenger approach to gravitational-wave astronomy, that seeks to optimally combine gravtitational wave and electromagnetic observations.  We highlight the importance of a multi-messenger approach for detecting gravitational wave sources, and also describe some ways in which joint gravitational wave and electromagnetic observations can improve the estimation of source parameters and better inform our understanding of the sources themselves -- thus enhancing their potential as probes of astrophysics and cosmology.

\end{abstract}


 
\acrodef{GW}[GW]{gravitational-wave}
\acrodef{BNS}[BNS]{binary neutron star}
\acrodef{CBC}[CBC]{compact binary coalescence}
\acrodef{HMNS}[HMNS]{hypermassive neutron star}
\acrodef{SGRB}[SGRB]{short-duration gamma-ray burst}
\acrodef{LGRB}[LGRB]{long-duration gamma-ray burst}
\acrodef{GRB}[GRB]{gamma-ray burst}
\acrodef{ET}[ET]{Einstein Telescope}
\acrodef{NS}[NS]{neutron star}
\acrodef{EM}[EM]{electromagnetic}
\acrodef{SNR}[SNR]{signal-to-noise ratio}
\acrodef{PDF}[PDF]{probability distribution function}
\acrodef{EOS}[EOS]{equation of state}
\acrodef{GWGC}[GWGC]{Gravitational Wave Galaxy Catalogue}
\acrodef{UNGC}[UNGC]{Updated Nearby Galaxy Catalog}
\acrodef{MMPF}[MMPF]{multi-messenger prior function}


\section{Introduction}\label{sec:intro}%
As we mark the centenary of General Relativity, the rapidly emerging field of gravitational wave astronomy stands on the threshold of a new era, with the global network of so-called `second generation' ground-based interferometers preparing to begin operations \citep{Virgo:2009,Harry:2010}.  The approximately ten-fold improvement in sensitivity of these detectors compared with the first generation LIGO and Virgo detectors \citep{2015CQGra..32g4001T} means that the first direct detections of gravitational waves (GWs) are eagerly anticipated within the next few years -- with the most likely sources expected to be the inspiral and merger of compact binary systems:  neutron star-neutron star (NS-NS) binaries; black hole-black hole (BH-BH) binaries or NS-BH binaries.   The challenges associated with detecting such sources, and exploiting them as probes of astrophysics and cosmology, have attracted considerable effort in recent years -- including thorough investigations of the GW detection rates that can be expected \citep{2010CQGra..27q3001A}, the most likely electromagnetic (EM) counterparts that will accompany them \citep{2012ApJ...746...48M} and the efficiency with which those EM counterparts could be detected \citep{2013ApJ...767..124N,2015arXiv150307869C}.  What is clear from this work is that significant benefits will derive from {\em joint\/} observations of both GW signals and their EM counterparts -- giving rise to a new observational field that has been termed `multi-messenger' astronomy.

The benefits of a multi-messenger approach are manifest for both the detection and characterisation of GW sources.  Perhaps the most immediate benefit is a precise identification of the source sky location -- alleviating the relatively poor sky localisation that is possible from ground-based GW observations alone \citep{Wen2008JPhCS.122a2038W,Wen2010PhRvD..81h2001W,2011ApJ...739...99N,Schutz2011CQGra..28l5023S,2011CQGra..28j5021F,2014PhRvD..89h4060S}.  Another clear benefit derives from the potential identification of an EM `trigger', or associated emission, that may streamline and enhance GW searches by reducing the number of free parameters, or at least significantly restrict the range of their values that must be explored \citep{2004CQGra..21S.765M,2008ApJ...683L..45A}.  Equally, a multi-messenger approach may involve using GW `triggers' to prompt targetted EM searches for the counterpart of the GW source \citep{2012A&A...541A.155A,2014ApJS..211....7A} -- a successful outcome of which would indeed both identify a precise sky location and permit follow-up observation of the source's host galaxy,  allowing for example measurement of its redshift.  More generally, joint GW-EM observations may in turn lead to a better and more robust inference of the source parameters, thus opening the way to using GW sources more effectively as astrophysical probes.

There are very significant observational and computational challenges associated with these multi-messenger approaches, however.  From a GW point of view, there is the challenge of carrying out a sufficiently rapid real-time analysis of the interferometer data -- thus allowing meaningful information to be supplied to the EM community quickly enough to permit follow-up observations of any prompt EM emission associated with the GW event \citep{2012A&A...541A.155A,2012A&A...539A.124L,2012ApJS..203...28E,2014ApJS..211....7A}.   To act upon the results of such an analysis then presents major challenges from an EM point of view:  e.g. understanding what {\em are\/} the EM counterparts themselves, and what are their signatures across the EM spectrum, and then searching efficiently for those signatures -- the latter task rendered difficult in view of the poor sky localisation provided by the GW triggers, particularly during the first few years of operation of the second generation network \citep{2014ApJ...795..105S}.

In this paper we will consider some of these multi-messenger issues in more detail for the particular case of NS-NS or NS-BH compact binary coalescences (CBCs). We will very briefly summarise our current understanding of the expected EM counterparts of these sources and highlight some of the observational issues this raises for their detection through rapid follow-up observations across the EM spectrum.  We will also briefly discuss how GW searches can provide suitable triggers to the EM community on an appropriately short timescale.  Finally, we consider the astrophysical potential for multi-messenger observations of CBCs -- highlighting one specific cosmological question that may be addressed in the fairly near future with these data.

Of course a multi-messenger approach is potentially crucial for {\em all\/} GW sources -- not just CBCs -- and may become increasingly important as the `detection era' of gravitational-wave astronomy unfolds.  For example, looking ahead to possible third generation ground-based detectors such as the proposed European Einstein Telescope \citep{2010CQGra..27s4002P} there are excellent prospects for multi-messenger science (involving not just GW and EM observations but also cosmic rays and neutrinos, see for example \citep{2014PhRvD..90j2002A}) with a wide range of astrophysical sources.  The interested reader can find more details in e.g. \cite{2011GReGr..43..437C,2013CQGra..30s3002A} but we do not consider the broader context for future multi-messenger astronomy any further in this paper.

\vspace*{-1mm}
\section{Joint detection of GW and EM emission:  constraints from the event timescale}

To make best use of a multi-messenger approach to observing GW sources requires both a solid theoretical understanding of the signatures produced by the EM counterpart of the source, together with the ability to observe these signatures straightforwardly.  Consider the example of a NS-NS or NS-BH CBC merger event, for which the likely counterpart is thought to be a short-duration gamma ray burst (sGRB) \citep{1986ApJ...308L..43P,2005Natur.438..988B}. \cite{2012ApJ...746...48M} have addressed in detail the question of what might be the most promising electromagnetic counterpart of an sGRB, and they identify four key features (which they term `cardinal virtues') that such a counterpart should possess.  It should:
\begin{enumerate}
\item{be detectable with present or upcoming telescope facilities, provided a reasonable allocation of resources}
\item{accompany a high fraction of GW events}
\item{be unambiguously identifiable, such that it can be distinguished from other astrophysical transients}
\item{allow for a determination of $\sim$arcsecond sky position}
\end{enumerate}

We will return to virtues 2, 3 and 4 in the next section, but the physical nature of a sGRB event makes satisfying virtue 1 already a challenge -- particularly when one is considering EM searches prompted by a GW trigger -- due to the very short timescales involved.   According to the standard picture, in the first one or two seconds  immediately after the merger the sGRB results from rapid accretion onto a centrifugally supported disk that surrounds the merged object; this accretion powers a highly relativistic jet that, due to beaming, is only seen as a sGRB within a narrow half-opening angle of a few degrees \citep{2006ApJ...653..468B,2006ApJ...650..261S}.  Interaction of the jet with the interstellar medium produces prompt afterglow emission that is visible in X-rays on a timescale of seconds to minutes after the merger, and in the optical on a timescale of hours to days -- but again only within a narrow solid angle around the jet.  On the other hand it is believed that the event will also emit fainter isotropic radiation, in the optical and infra-red, on a timescale of days after the merger; this emission is caused by the decay of  heavy elements synthesised by the merger ejecta as it interacts with the surrounding environment, and is referred to as a `kilonova' \citep{1998ApJ...507L..59L,2005astro.ph.10256K,2010MNRAS.406.2650M,2015MNRAS.450.1777K}. We will see in the next section that kilonovae may be key to the successful identification of an EM counterpart for many CBC events.

Thus, as is summarised in e.g.  \cite{2014ApJ...795..105S} and \cite{2015arXiv150803634S},  the nature of the EM counterparts of a NS-NS or NS-BH merger, and their rapid evolution in time, requires that the GW trigger be generated as quickly as possible -- and ideally on a timescale of seconds to (at most) hours -- particularly if the event afterglow is to be observed in X-rays.   

Such a {\em low latency\/}, template-based data analysis pipeline has been developed for searching for CBC signals  with the ground-based interferometer network -- and the need for a very rapid trigger generation is one of several key factors driving this search. (For more details see \cite{2012ApJ...748..136C} ).  In the sixth LIGO science run and third Virgo science run,  the inspiral search pipeline Multi-Band Template Analysis (MBTA) \citep{2008CQGra..25d5001B} was able to achieve trigger-generation latencies of 2--5 minutes (See e.g. \citep{2012A&A...541A.155A}).  Interesting co-incident triggers identified by the MBTA were submitted to the Gravitational-wave Candidate Event Database (GraCEDb) \footnote{https://gracedb.ligo.org/}.  However, with the advanced detector network, given that sources will spend up to 10 times longer within the detection band, the  low-latency trigger generation of inspiral signals will become even more challenging.  

The  slow evolution in frequency of an inspiral signal allows one to reduce the sampling rate of both the data and templates.   The  highly similarity between neighbouring templates in the filter banks allows one effectively to reduce the number of filters or to use  so-called `parallel  infinite impulse response'  filters.   We refer the reader to the LLOID (Low Latency Online Inspiral Detection) \citep{2012ApJ...748..136C} and SPIIR (Summed Parallel Infinite Impulse Response) \citep{2012PhRvD..85j2002L,2012PhRvD..86b4012H,2012CQGra..29w5018L} pipelines for details of the implementation of these efforts.   

\section{Joint detection of GW and EM emission:  identifying the source location}

From analysis of the GW interferometer network data alone, an approximate sky position for a CBC source can be obtained by exploiting the difference in arrival times of the GW signals at the different detector locations \citep{2013arXiv1304.0670L}.  In the first few years of operation of the advanced detectors the typical `error box' derived in this manner will, unfortunately, have an area of several hundred square degrees \citep{2014ApJ...795..105S}.  The situation will improve substantially in later years as more interferometers are added to the global network, and by the early 2020s -- with the anticipated inclusion of the proposed LIGO India detector -- almost $50\%$ of binary NS merger systems detected by the network are expected to be localised to within 20 square degrees on the sky.  Looking even further ahead, recent work \citep{2013CQGra..30o5004R,2015CQGra..32j5010H} investigating the optimal sites of future, third generation, detectors has applied various metrics to define and quantify the performance of a given network configuration -- including the ability of the network to localise the position of a typical CBC source on the sky.  However, a fundamental limitation is set by the scale of the Earth, and the maximum difference which this imposes on the time of arrival of GWs at well separated detectors.  Thus, for {\em any\/} ground-based detector network, present or planned, we can expect that GW data alone will locate many sources to a precision of only tens or even hundreds of square degrees.  Clearly, then, identifying a unique EM counterpart would be a crucial benefit of a multi-messenger approach as it can provide a much more precise sky location for the GW source.

So how might such an EM counterpart be identified, given such relatively poor sky localisation information, in the context of the `cardinal virtues' highlighted above?  As discussed in \cite{2012ApJ...746...48M}, sGRBs are not in themselves likely candidates since the beaming of their relativistic jets will mean that most merger events will not be seen as a sGRB at the Earth -- even if a sGRB does indeed `accompany a high fraction of events'.  Similarly the prompt X-ray and optical afterglow emission associated with the sGRB may also not be seen at the Earth -- although if such an afterglow {\em were\/} observed through rapid follow-up observations with a high-energy burst monitor satellite such as Swift, then this can in principle determine a very precise ($\sim 10$ arcseconds) sky location for the GW source \citep{2012ApJS..203...28E}.  As concluded in \cite{2012ApJ...746...48M}, therefore, kilonovae are perhaps the most promising candidates for the EM counterpart of a NS-NS or NS-BH merger event since their isotropic emission gives them the potential to satisfy all four cardinal virtues.  On the other hand their comparative faintness and their limited duration still suggests that their detection will be very challenging, involving rapid, wide-field EM follow-up campaigns. In fact to date only two candidate kilonovae events have been observed \citep{2013Natur.500..547T,2015NatCo...6E7323Y}.

Whatever the nature of its EM counterpart, a key step towards its study is the (hopefully  unique) identification of the host galaxy of the GW source.  To this end we can develop a general framework for combining GW and EM information under the straightforward assumption that there are common parameters -- such as sky location and distance -- among the observations made of a GW source, its EM counterpart  and their host galaxy.  Such a multi-messenger approach has already proven fruitful even in the {\em absence\/} of any observed GWs.  For example, searches for GWs associated with the gamma-ray bursts  GRB070201 \citep{2008ApJ...681.1419A} and GRB051103 \citep{2012ApJ...755....2A}  have ruled out the possibility that their progenitors were binary neutron stars (NS)  sources in Andromeda galaxy (M31) and M81, respectively. 

One can consider the question, then, of how the detection efficiency of an EM counterpart might be improved by conducting wide-field EM follow-ups that make use of pre-existing galaxy catalogs.  This issue has been investigated in  \cite{2014ApJ...784....8H}; by taking into account the GW measurement error in both distance and sky location they estimated that an average of $\sim 500$ galaxies are located in a typical GW sky location error box for NS-NS mergers with Advanced LIGO ($\sim$20 deg$^2$), up to range of 200 Mpc. The authors then found that the use of a complete reference galaxy catalog could improve the probability of successful identification of the host galaxy by $\sim 10-300 \%$ (depending on the telescope field of view) relative to follow-up strategies that do not utilize such a catalog.

Of course a more complete treatment ideally should involve a comprehensive end-to-end simulation that takes fully into account the characteristics of the global GW network, the spatial distribution and event rate of CBC events, the multi-wavelength light curves of their EM counterparts (together with the event rate and lightcurves of a range of possible `false positives' with which these counterparts might be confused, bearing in mind virtue 3 above) and the sensitivies and observing strategies adopted by the multi-wavelenth EM telescopes used to search for them.  Much work remains to be done in this area, both on more detailed theoretical modelling of the EM counterparts themselves (e.g. kilonovae) and on investigating the impact and efficacy of different follow-up observing strategies.  Some excellent progress has already been made, however, particularly in \cite{2013ApJ...767..124N} where the authors point out that:
\begin{itemize}
\item dedicated 1m class optical and near infra-red telescopes with very wide-field cameras are well-positioned for the challenge of finding EM counterparts 
\item a comprehensive catalog, out to z $\sim$ 0.1, of foreground stellar sources, background active galactic nuclei and potential host galaxies in the local universe -- and probing deeply (i.e. approximately 2 magnitudes deeper than the expected magnitudes of the EM counterparts) -- is needed to assist identification of the EM counterparts amidst the larger numbers of false positives.
\end{itemize}
 
To date no such all sky galaxy catalog is available.  Among the many existing galaxy catalogs, the Gravitational Wave Galaxy Catalog \citep{White2011CQGra..28h5016W}(GWGC;(White et al. 2011) has been specifically compiled for current follow-up searches of optical counterparts from GW triggers -- although the authors acknowledged that the catalog is increasingly incomplete at larger distances.  \cite{2010PhRvD..82j2002N} proposed a ranking statistic to identify the most likely GW host galaxy (drawn from the GWGC) based on galaxy distance and luminosity and the sky position error box. This ranking method has been adopted in the design of an EM follow-up pipeline  ( such as \cite{2013ApJS..209...24N})  and follow-up observations (such as \cite{2014ApJS..211....7A,2012ApJ...759...22K,AndoRevModPhys.85.1401}).

 A novel Bayesian approach to identifying the GW host galaxy, incorporating astrophysical information has been proposed in \cite{2014ApJ...795...43F}. A merit of this approach is that both the rank and the posterior probability of a galaxy hosting the GW source are estimated with the help of  GW-EM relation models.   Using a simulated population of CBCs the authors found that (i) about 8\%, 4\%, and 3\% of injections had 50\%, 90\%, and 99\% respectively of the probability to be included in the top 10 ranked galaxies in the GWGC, and (ii) the first ranked galaxy had a 50\% probability of being the true GW host galaxy in about 4\% of injections. These results are dominated by the GWGC distance cut of 100 Mpc, compared with the expected reach of $\sim$200 Mpc for Advanced LIGO and Advanced Virgo at design sensitivity.  The method could easily be extended and applied to deeper galaxy catalogs, however.

A comprehensive, complete  all sky catalog of {\em all\/} objects (galaxies, stellar sources and so on) may not be essential  for a single GW-EM joint observation, since the  GW sky map will only cover a small fraction  of the whole sky.   A ``galaxy survey on the fly" approach has been proposed to reduce the required effort and thus enhance the immediate availability of useful catalogs \citep{2015ApJ...801L...1B}; in this approach a rapid galaxy survey using 1-2 m class telescopes can efficiently catalog those galaxies covered by one GW detection volume within a short period of time.   This rapidly compiled galaxy catalog could then very quickly be provided to other telescopes, to aid further electromagnetic follow-up observations of e.g. kilonovae, as well as other sources.

Whatever detailed strategies are employed in the future, it is clear that to identify the common GW and EM source one needs multi-wavelength  observations capable of rapidly and efficiently targeting multiple objects.  A community of `multi-messenger' astronomers -- including the LIGO and Virgo collaborations and a significant and growing number of telescopes across the world -- is now being assembled for exactly this purpose$^1$\footnote{$ 1.  \, \rm https://gw-astronomy.org/wiki/viewauth/LV\_EM/$}.

\section{Beyond detection}

The promise of multi-messenger astronomy -- and in particular of joint GW and EM observations of CBCs and their EM counterparts -- lies not just in enhancing the prospects for detecting these events but also in improving the estimation of their parameters, and ultimately their use as astrophysical and cosmological probes. 

Consider again, for example, the case of a CBC event. In the absence of any spin, the inspiral phase of a CBC in a circular orbit has a gravitational waveform that depends on the following nine parameters: four intrinsic parameters -- the two masses and two intrinsic constants of integration, which define the phase evolution of the waveform; five extrinsic parameters -- luminosity distance, sky location, inclination angle (the angle between the orbital angular momentum and the line of sight) and GW polarization.  All nine parameters govern the amplitude of the signal that each GW detector should `see'.  So we see immediately that the identification of a unique EM counterpart will begin to reduce the dimensionality and complexity of this parameter space, providing tight constraints on the sky location and the luminosity distance (e.g. via a measured redshift for the host galaxy -- although deriving a luminosity distance directly from this will be complicated by the effects of galaxy peculiar velocities).  Knowledge of the luminosity distance from an EM counterpart will also help to break parameter degneracies that may exist for the GW source, such as that between distance and inclination angle \citep{2010ApJ...725..496N}.  

More generally, joint GW-EM observations should allow for better characterization of the signal progenitor, and a richer interpretation of the results of GW searches -- offering, for example, greater insight at the population level into the relationship between the GW source, its EM counterpart(s) and their progenitor environment.  An excellent example of such an approach is the relationship between CBCs and sGRBs that we have discussed in detail in this paper, but as the detection era unfolds no doubt other similar examples will emerge.  The insights derived from these joint analyses may improve our understanding of the evolutionary processes that led to the GW event, and thus perhaps provide better prior information about the likely sites (e.g. in terms of host galaxy morphology, colour, metallicity etc) of future events.   Moreover, as our understanding of these deep, underlying astrophysical relationships improves, new data analysis methods will be built to explore them -- combining not just GW and EM observations but also data from neutrino and cosmic-ray telescopes \citep{2013JCAP...06..008A,2014PhRvD..90j2002A}.  

Underpinning this joint approach at all stages is the fundamental assumption that these multi-messenger phenomena all emanate from the same object, and obey physical relationships that share some common parameters.  Bayesian inference then provides a natural framework in which to incorporate multi-messenger astrophysical information and optimally estimate those parameters. The method introduced in \cite{2014ApJ...795...43F,2015ASSP...40...35F} explores some specific examples of this framework in action; these include reducing the area of sky over which astronomical telescopes must search for an EM counterpart, improving the inference on the inclination angle of a NS-NS binary and estimating the luminosity of a sGRB.  

What specific astrophysical questions will be tackled via a multi-messenger approach in the future?   We end by briefly considering one such example that is anticipated for the advanced detector era.

It was recognised nearly thirty years ago by Schutz (1986) \cite{Schutz1986Natur.323..310S} that CBCs have the potential to be precise luminosity distance indicators via measurement of the time-evolution of their amplitude, frequency and frequency derivative during the inspiral and merger phase.  This has given rise to the idea of so-called gravitational-wave {\em standard sirens\/}, by analogy with the {\em standard candles\/} of EM astronomy -- although the CBCs do not require any assumption to be made about their intrinsic `luminosity' (which essentially depends on the masses of the binary stars) as this can be inferred directly from the observed data at the same time as the luminosity distance.  In more recent years there has, therefore, been much interest in the potential use of standard sirens as cosmological probes, via calibration of the luminosity distance redshift relation.  

The expected reach of advanced detectors will be too shallow to permit exploration of dark energy models and the accelerated expansion of the Universe -- although such models could certainly be investigated by third generation detectors such as the Einstein Telescope \citep{2010CQGra..27u5006S,2011PhRvD..83b3005Z} or spaceborne missions such as eLISA \citep{2005ApJ...629...15H,2007ApJ...668L.143D,2012CQGra..29l4016A}.  However, a realistic target for the upcoming global network of advanced detectors is measurement of the Hubble constant, $H_0$, using standard sirens.  Recently  \cite{2013arXiv1307.2638N} have investigated the efficacy of such a measurement,  and conclude that a precision of about $1\%$ on $H_0$ is possible from observations of about 30 NS-NS mergers within a few hundred Mpc -- using the anticipated future global intereferometer network that includes KAGRA and LIGO India.  Such a precise value would certainly be competitive with the EM results expected on a similar timescale from e.g. the James Webb Space Telescope \citep{2010ARA&A..48..673F}, and in any case would be an extremely useful adjoint to a purely EM determination of $H_0$ using the traditional cosmic distance ladder since it would be subject to a completely different set of systematic uncertainties.

A standard siren measurement of $H_0$ will present a major multi-messenger challenge for several reasons.  Firstly, to estimate the Hubble constant of course requires comparison of distance with redshift, and the latter will not generally be measurable from GW data alone (but see also below for discussion of some interesting alternative approaches). Indeed the luminosity distance estimates for the sirens will in any case be degenerate with redshift because of the mass-redshift degeneracy in post-Newtonian CBC waveforms.  By measuring the redshift of the siren's host galaxy the degeneracy can immediately be broken.  However, this measurement of course first requires the prompt observation of an EM counterpart and the unique identification of the host galaxy -- steps which will be subject to all of the multi-messenger issues discussed in the previous section. 

Notwithstanding these potential difficulties, and their resulting impact on the final error budget for $H_0$ (for which the the estimate of $1\%$ in \cite{2013arXiv1307.2638N} may therefore be somewhat too optmistic), the prospect of a gravitationally-calibrated value of the Hubble constant is nevertheless extremely exciting -- and is likely to be one of the main targets for gravitational-wave astronomy over the next decade.

It is interesting to note that some other approaches to breaking the mass-redshift degeneracy, and/or determining redshifts (and hence cosmological parameters) from GW data alone, have been proposed.  For example \cite{2012PhRvD..85b3535T} assume that there exists a universal (rest frame) mass distribution for NSs at different redshifts and by comparing the measured (redshifted) mass distribution of NSs with the local mass distribution show that one could infer statistically the redshifts of the sources and hence derive indirectly the value of $H_0$. Their results show that, in this way, second generation interferometers should be able to infer the Hubble constant with $\sim 10\%$ accuracy from about 100 events. Their analysis is extended in \cite{2012PhRvD..86b3502T} to consider the cosmological potential of GW-only observations with third generation ground-based detectors.

In a similar manner, various authors \cite{2008PhRvD..77d3512M}, \cite{2011ApJ...732...82P} have proposed that the identification of the host galaxy -- and thus the determination of its redshift -- may be carried out {\em statistically\/}, using prior information about the spatial distribution of the galaxies in the sky localisation error box provided by the GW data alone.  Using this formalism del Pozzo (2014) \cite{2014JPhCS.484a2030D} suggests that the Hubble constant could be determined to a few percent from observations of about 50 sirens with the advanced detector network.

A third, highly promising, possibility for constraining cosmological models using GW observations alone has been proposed by Messenger \& Read (2012) \cite{2012PhRvL.108i1101M}, exploiting the effect of tidal deformations on NS-NS binary systems during the inspiral that provide additional contributions to the phase evolution of their gravitational waveforms. Recently their approach has been studied further, and is predicted to be capable of determining redshifts to a precision of $10 -- 20\%$ for GW sources in the local Universe observed with the Einstein Telescope \cite{2014PhRvX...4d1004M}.   The potential of this approach for determining cosmological parameters with Einstein Telescope observations has been further explored in \cite{2015arXiv150606590D}.

While these various methods suggests the intriguing possibility, therefore, of measuring $H_0$ -- and indeed other cosmological parameters -- without using {\em any\/} EM observations directly, the generally lower precision of these estimate does, nonetheless, underline that a multi-messenger approach will usually be more effective than a GW-only analysis.

\section{Summary}

In this paper we have discussed some of the advantages -- both for the detection of GW sources and also for the estimation of their parameters and their astrophysical exploitation -- of a multi-messenger approach that seeks to combine optimally GW and EM observations.  Focussing mainly on the inspiral and merger of NS-NS and NS-BH binaries, which are believed to be the progenitors of short duration GRBs, we have also highlighted some of the important observational challenges that need to be overcome in order that a multi-messenger approach may be fully exploited.  These challenges present significant logistical and computational constraints for the analysis of data from the ground-based network of advanced GW interferometers.  This is because of the very short timescale (from seconds to hours to days at most) associated with the EM counterparts of these events and the relatively poor sky localisation provided by the GW data alone -- which will significantly complicate the search for a unique EM counterpart. Nevertheless substantial progress has already been made towards establishing a community of `multi-messenger' astronomers, working closely together towards the goal of making joint GW-EM observations, and there are excellent prospects for this emerging new field over the next decade.

Looking further ahead, the potential of multi-messenger astronomy would appear to grow even stronger with the possible advent of third generation ground-based interferometers such as the Einstein Telescope and the possible launch of a spaceborne GW detector such as eLISA.  Although we did not discuss these future missions and projects in this paper, they should offer exciting science possibilities such as probing the equation of state and internal structure of neutron stars, constraining models of core-collapse supernovae and mapping the detailed structure of spacetime around black holes.  As general relativity enters its second century, the prospects for testing Einstein's theory under extreme cosmic conditions using multi-messenger data look very bright.
\acknowledgments
 X.F. acknowledges financial support from the National Natural Science Foundation of China (grant No. 11303009). M.H. acknowledges the hospitality and financial support provided by the Kavli Institute for Theoretical Physics in Beijing.



\bibliographystyle{apj}
\bibliography{Bibliography_2}

\end{document}